# A Framework For Analysis and Comparison of Dynamic Malware Analysis Tools

Waqas Aman

Norwegian Information Security Laboratory, Gjøvik University College, Norway

*ABSTRACT*

*Malware writers have employed various obfuscation and polymorphism techniques to thwart static analysis approaches and bypassing antivirus tools. Dynamic analysis techniques, however, have essentially overcome these deceits by observing the actual behaviour of the code execution. In this regard, various methods, techniques and tools have been proposed. However, because of the diverse concepts and strategies used in the implementation of these methods and tools, security researchers and malware analysts find it difficult to select the required optimum tool to investigate the behaviour of a malware and to contain the associated risk for their study. Focusing on two dynamic analysis techniques: Function Call monitoring and Information Flow Tracking, this paper presents a comparison framework for dynamic malware analysis tools. The framework will assist the researchers and analysts to recognize the tool's implementation strategy, analysis approach, system-wide analysis support and its overall handling of binaries, helping them to select a suitable and effective one for their study and analysis.*

## 1. INTRODUCTION

The internet has evolved exponentially since it was first launched in 1960. It has grown increasingly from a four node communication model to a highly interconnected and sophisticated network providing more and more services inheriting boundless benefits. The marketplace of products and services are broadening on a global level, new ways of social interactions are established and online banking has given new dimensions to banking operations. However, these benefits have also motivated the people with malicious intents who are looking for *any* opportunities to fulfill their objectives. Malware is one of the key mediums to furnish these malicious intents.

Legitimate users are protected from such malicious codes by the security vendors who provide them a software, antivirus, which identify and analyze these codes and alert the user accordingly. Typically, antivirus is equipped with a signature database which is used in the matching process to identify potential known or common threats. Malware analyst obtains a piece of suspected code and analyzes it to find whether it is harmful or not. When a threat in a code is confirmed, the analyst looks for a particular pattern of the threat and develops a signature for that code (malware) and is added to the signature database to confront a particular malware in the future. This manual process though seems trivial and does the job but is subjected to time consumption and errors as there exist a number of variant of the same code. Statistics shows that anti-malware vendors are experiencing thousands of malicious codes each day. McAfee, for instance, received more than 20,000 malware samples of mobile alone in 2012 [1]. According to NQ Mobile, mobile malwares alone has grown up to 163% and has increased to more than 65000 in 2012 [2]. Hence there is a need to automate the process to support quick, effective and timely analysis of such samples.

Typically, two methods are used to perform malware analysis; Static and Dynamic analysis. The distinguishing ground between the two techniques is that the dynamic analysis observes a malicious behavior while a sample code is executed whereas; the static approach does not execute the code. Though, they differ in the techniques used, a number of methods and tools were proposed to serve the same objectives of malware analysis. In this paper, two most common techniques used in dynamic malware analysis: Function call monitoring and Information Flow Tracking (IFT) will be conversed. Understanding these two techniques gives a general and holistic picture of a malware executable. Functional Call Monitoring aims to investigate the functional behavior and IFT provides and understanding of what data a malicious binary is interested in during execution. Thus, provides the analyst a fair idea of: for *what reasons* does a malware *misbehave*. However, analyzing a malware has turned into a crucial as well as an essential skill for security professionals and forensics investigators. Malware analysis not only enables an analyst to realize the malevolent code objectives but also gives an understanding of the evolving trend of a malware thus providing an analyst a tool to enrich the detection techniques. Due to the diversity of concept and strategies used in the implementation of these methods and tools, security researchers and analyst finds it difficult to select an appropriate tool for their investigations. Therefore, a framework is presented that aims to provide a platform to identify and understand the techniques that various tools offer to an analyst and the strategies they implement in accomplishing dynamic malware analysis thus helping them to select an optimum method or tool for their investigations.

The rest of the article is organized as follow: In section 2, an understanding of malware analysis will be highlighted, and preliminaries of Function Call monitoring and Information Flow Tracking will be conversed. In section 3, the proposed framework is presented. A summary of selected tools to be analysed and compared will be briefly described in section 4. Section 5, will provide a comparative analysis of the discussed tools and finally, the paper will be concluded in section 6.

## 2. MALWARE ANALYSIS

Malware is a piece of software or program that deliberately fulfill the malicious attempts of an attacker. Malware comes in many forms with different objectives. The common terms used to classify different types of malwares are Trojan, worm and virus, etc. [3]. Traditionally, they were developed either for fun factors to show one's capabilities or for highlighting weaknesses within a system. However, today, these motivations have gone to the highest level of treachery. Now we can see a spectrum of motivation ranging from personal to national level interest and a whole new underground economy is based on malwares these days [4]. Stuxnet [5] is one of the latest pictures of such motivations. Malwares are propagated using numerous infection vectors such as exploiting vulnerability on a client system, through an open or vulnerable network service, using removable devices [5, 6] or through social engineering. To combat against malwares, systems are now equipped with antivirus programs. Most of the antivirus programs consist of a scanner and a signature database. The scanner matches file on the user system and matches them against the available signatures. An alert is generated and the user is informed when a match is found.

Typically, two standard approaches are used to analyze the behavior of a malicious program. Dynamic analysis is a set of methods that are used to understand the behavior of a program during its execution while static analysis is used to investigate a program without executing it. The following subsections will elaborate static analysis and the shortcomings associated with it.

## 2.1. Static Analysis

Analyzing a program to observe its behavior by investigating it without execution is commonly known as static analysis. It can be performed in numerous ways depending upon the availability of the code and its representation. Static analysis can assist in evaluating the memory errors and can improve the correctness of a program execution if it source code is available [7, 8]. It can also be used to inspect a binary executable with different tools [9]. Static analysis can be prompted before or after dynamic analysis or can be done as a standalone procedure. Sometimes it is performed to see if the analysts have missed anything suspicious after the dynamic analysis. And is performed pre-dynamic analysis to analyze and understand the behavior prior to the code execution in a live environment.

### 2.1.1. Shortcomings of Static Analysis

Because source code of most of the programs is not readily available makes the static analysis approach harder to combat the malicious programs thus reduces its application. Analyzing binaries with static approaches have inherited complications and challenges. Some of the malwares with strong evasion and obfuscation techniques such as the presence of Opaque Constants make the disassembling of a binary executable ambiguous due to which the resulting code cannot be analyzed precisely [10]. Disassembling remains the essential part of the static analysis of binaries which can be easily made ambiguous by simple obfuscation measures. Such obfuscation techniques obscure the program flow, making the variables inaccessible and disable the tracking of values stored in a register. These limitations of the static approach motivate the development of analysis techniques that can overcome the mentioned code evasions and transformations and to analyze a malicious program accurately and reliably.

## 2.2. Dynamic Malware Analysis

Dynamic analysis is the process of executing malware in a monitored environment to observe its behaviors [11]. While a monitored program executes, it provide a detailed information readily such as URL accessed, files created and accessed, information transferred, registry keys creation and so forth. In this section, a detailed discussion will be made on two most common techniques used in dynamic analysis: Information Flow Tracking and Function Call Monitoring. Furthermore, a description of various approaches and concepts used to accomplish these techniques will also be elaborated. The objective of the underlying discussion is to understand two vital aspects of analyzing a malicious program i.e. their functional behavior and the data or information they are interested in while they execute.

### 2.2.1. Information Flow tracking

Information Flow Tracking (IFT) approaches are used to monitor programs from the perspective of how the program processes the data. During the analysis, suspected or interested data is tainted or labeled before it is processed and its propagation is then observed when the data is processed. Because of the tainting data, information flow tracking is also termed as Taint Analysis [12]. In [13], the author illustrates information flow tracking process with a simple example reviewed below in table 1. Considered a stack location, *z*, is an unsafe location. A received data, *x*, from the network is considered suspicious and is tainted. During processing, *x* is copied to another location, *y*. Now during program execution if the program jumps to the already tagged location z, an alarm will be raised indicating a malicious activity.

International Journal of Network Security & Its Applications (IJNSA), Vol.6, No.5, September 2014

Table1. Simple Information Flow Tracking

| Program Execution | Information Flow Tracking |
|---|---|
| ….. | Tag(z) = 1 Suspicious location is tagged unsafe |
| Received (&x); | Tag(x) = 1 Unsafe data tagged |
| y = x; | Tag(y) = Tag(x) |
| ….. | ….. |
| Jmp z; | Raise Alarm, as z is already tagged as unsafe & suspicious |

Some of the key concepts used in the process of IFT analysis are described below:

- *Tainted Sources and Destinations:* Taint sources are the components of an Information Flow Tracking system that labels interested or suspicious data in the system. They taint the data and give them as an input to the system. Tainted destinations, on the other hand, are the elements which respond to a tainted data introduced by the taint sources in a tailored mechanism. Responses can either be triggering another source or raising an alert.
- *Direct Tainted Data:* As the name implies, direct tainted data refers to labelling an operand directly. The common rule of data tainting is that when a tainted source value is moved or copied to another operand, the destination operand is also tainted [14]. However, an IFT system may contain more than one direct tainted operands involved in a single (or multiple) instruction(s) with distinct labels as depicted in table 2. A guideline should be maintained that direct how these tainted operand should be handled. For example, the most suspicious or interested operand can be monitored over the other or they can be analyzed in a combined manner under a common label.

Table 2. Direct Data Tainting

| Program Execution | Information Flow Tracking |
|---|---|
| y = x; | Tag(y) = Tag(x)  Direct tainted data with distinct tags |
| y = y + x; | Tag(y) = Tag(x) + Tag(x)  Direct tainted data with distinct tags in an arithmetic operation |

- *Tainting Addresses:* Pointer or address tainting is used when the information flow tracking system is used to observe non-control data exploits or analyzing sensitive information leakage and is done by generating an address using a tainted data [15]. When an address *A* is tainted, it is detected as a malicious activity if *A* is de-referenced during the program execution. Consider the example in table 3 below:

Table 3. Tainting Addresses

| Program Execution | Information Flow Tracking |
|---|---|
| x = 0; | Tag(y) = Tag(x)  Direct tainted data with distinct tags |
| A = &x; | Tainted Address A generated using tainted value X |
| A = 1; | Tainted Address A is de-referenced – Generate Alarm |

- *Control Dependencies:* Besides data and address tainting, information flow tracking can also be performed using control flow tainting. For instance, an instruction *X* is control dependent on instruction *B*, if *B* controls the execution of *A* [15]. An illustration of control-flow tainting is provided in table 4. It implies that an address or data can only be tracked if another tainted operand on which its execution is dependent is also processed by the system.

Table 4. Control Flow Tracking

| Program Execution | Information Flow Tracking |
|---|---|
| If (A == 0) { B == 1; } | T ag(B) ⇔ T ag(A) = 0 |

- *Taint Policy:* Taint policy is a set of rules that direct how a taint analysis should be carried out within the information flow tracking system and covers three aspects of taint analysis; taint Introduction, Propagation and Checking [15]. Taint introduction rules direct how taint should be introduced to the system, propagation specifies how taints statuses are to be updated for data derived from tainted or untainted operands and checking is performed to validate the taint status and the corresponding action to be taken.
- *Over and Under-tainting:* These are the two types of errors that can be experienced during information flow tracking analysis [15]. Overtainiting occurs when the system marks a data as tainted when it is not derived from a tainted source. Undertainting occurs when the analysis process misses a value to be tainted in the information flow tainting.

## 2.3. Function Call Monitoring

A function is a subroutine or a code block in a program that execute specific operations. They are used whenever a code or task is repeated. This property makes functions reusable and easy to maintain if changes are required for a particular procedure. Functions are designed in an abstract manner with more semantic oriented approach rather than focusing on the implementation details which make it adaptable for many platforms. From the analysis perspective, such abstraction provides an understanding of the overall behavior of a program. A function call is a simple command which invokes a function by calling its name. Functions are usually monitored by intercepting the calls made to it.

The process of *hooking* is used to capture the function calls. The monitored program is operated in such a way that whenever a call to a function is made, the hook function is also invoked. Hook is a function itself which implements the analysis procedure and performs tasks like logging target program execution, observes intermediary function calls and analyzing various inputs and outputs.

The get an understanding of dynamic analysis in the context of functions and function calls, the under mention terminologies are explained to recognize how functions and function calls are triggered, the tasks they perform and the level they execute on.

- *Application Programming Interface:* API is a set of programming specifications that perform various functionalities. Operating Systems provide various API libraries of object classes, functions, data structures, and variables that are grouped together to accomplish common functions. They are offered at different layers of operating system. Windows Cryptography API [16], for instance, focuses on the provision of constants, command-line tools and functions, etc. required by the cryptography and certificate services used in a Window operating system.
- *System Calls:* Operating Systems provides two modes of program execution: User and Kernel mode. User programs such as word processing, imaging applications and browsing, etc. are executed in the user mode while the operating system runs its own programs in kernel mode. User mode processes don't have a direct access to the system level which can only be accessed by kernel mode process [17]. However, in order to perform various tasks such as creating a file on disk, a user mode process have to call a system level process. In operating systems, this is facilitated through special APIs called *System Calls*. The user mode switches to the kernel mode when a system call is triggered and the request can be executed then in a privileged mode at system level. Although there are some kernel level malwares which gain privileges to extract sensitive information such as Rootkits [18] however, most of the malwares infects user mode processes and make their way to the privileged kernel mode. Another way of invoking system calls in Windows operating systems are the Windows Native APIs. However, they differ from the usual managed APIs (discussed earlier) which remain the same in every operating while the native APIs varies from one service pack to another [19]. Some user mode processes use native APIs as a communication medium between higher level APIs and the system calls. Generally, a legitimate user mode application connects to the kernel level (system calls) through Windows API however; a malware may bypass the managed API layer triggering the native APIs directly to execute malicious code at the kernel level. As a result, the dynamic analysis tools which use API hooking to observe malicious behavior skip these malwares unnoticed.

- *Hooking:* API hooking enables the analyst to track specific behavioral elements of the program execution such as intermediary calls and parameters used in a function. Hooking to native interfaces can provide a truer and detail behavioral analysis of a malicious function and provide an insight to reveal suspicious activities bypassing the managed APIs. From the context of implementation, hooking can be performed in a variety of ways to circumvent malicious code from its execution. The simple way of hooking onto a function is to insert the hook function appropriately into the code, provided that the source code is readily available. [20] provides a set of function in GNU Compiler Collection (GCC) to support function hooking using flags during code compilation. Binary analysis mechanisms, for instance, [18], can be used to detect and analyze functional behavior, if binary executable are in hand.
  Windows operating systems support function hooking using the Detours Library [21]. It implements function hooking whenever a monitored function is called by diverting the control flow to a hook. When the hook function is initialized for analysis, it then calls in monitored function itself. Another way to achieve function hooking is debugging. It allows the analyst to insert breakpoints into the code at defined functions to observe its behavior. A number of techniques to debug Windows programs can be found at [22]. Breakpoints give a full control at the debugger level to access various process components for analysis.
- *The Trace:* The resulting product of function hooking is termed as a function trace. It is more like an analysis report containing rich information such as the parameters accessed and processed and the extended functions called by the monitored function. However, a raw trace file may be very complicated to understand and analyze. Different tools and methods are available which sort the trace information in a required semantic representation. Malicious and legitimate behavior of a function can be reviewed by using a graph where the differences between the graphs can show the presence of malicious activity [23].

## 3. PROPOSED FRAMEWORK

This section details the key concepts and strategies based on which the proposed framework is developed. They are necessary to understand or analyse dynamic malware analysis tools.

### 3.1. Analysis Level and Environment

Analysis system can be implemented at different level of an operating system. Its design assessment depends upon the type of malwares under investigation and the level of information extraction. Kernel mode analysis can extract rich information such as the systems calls and the intermediary Native APIs called. As a privileged mode operation, kernel mode analysis can hide its execution from the user level malwares [18]. On the other hand, user mode analysis can provide a detail description of the managed API calls and the invoked functions. However, it cannot retrieve the information on the invoked native APIs and corresponding system calls.

Analyzing malwares in native environment can lead to infections therefore analysis is done in a protected environment to analyze the monitored programs thoroughly without any risk of infection. Analysis can be performed in an emulated environment which enables an analyst to execute a binary program without being infected. The process is usually carried in a Sandbox which allows a program to be executed in an environment where all systems components such as I/O devices, CPU, memory and network services are emulated [24]. Similar kind of approach is the Virtual Machine Monitoring (VMM) where analysis is carried out in a virtual, isolated machine that runs on a real (host) system.

### 3.2. Monitored Process

A vital property of an analysis tool is its capacity to monitor to multiple processes at the same time. Most of the analysis techniques observe a single process however; malwares usually tend to be dependent on other actions (processes) or conditions to be triggered such as logic bombs [3]. Monitoring multiple processes enables an analyst to perform an in-depth investigation of activities and execution paths executed by a malware. Thus, in order to evaluate an analysis tool we can breakdown the analysis to single, spawned (where a monitored process activates a child process) and system-wide process monitoring.

### 3.3. Analysis Method

Analysis techniques refer to the investigation methods which are used or available in a particular tool. The two important analysis techniques focused this paper are the function call monitoring and IFT. However, the implementation or analysis level of a tool restricts an analyst to investigate a malware by using only few analysis methods. For instance, if an analysis tool is implemented in user mode, an analyst cannot observe or monitor the system calls or the intermediary native APIs. Thus as a thumb rule, tools incorporating more analysis methods and options provides a better analysis environment and produce positive as well as detailed results.

### 3.4. Unpacking the Binaries

Packing is a process in which a binary executable is packed in a transformed version of a program, which has a different syntax but have the same semantics. This technique is now used by most of the malware writers to hide the malware semantics from an analysis tool using different obfuscation and encryption measures. An unpacking module is appended with the packed malware, which is activated in the memory of the victim's system thus leaving no trace on the secondary memory. Though, analysis can be performed after the malware is unpacked but it is more advantageous for the analysts to perform analysis while it is packed as it enables them to understand the unpacking techniques used. Various techniques such as Reconstruction [25], Write xor Execute (W$\oplus$X) Policy [26] or detecting packer with packer' signature database is used to handle packed binaries.

## 4. TOOLS AND TECHNIQUES

This section presents a summary of eight most cited tools and techniques that incorporate Function Call monitoring and IFT to approach dynamic malware analysis. Furthermore, a brief overview will be provided on how individual tools address the features and strategies discussed in section 3.

*Panorama* [27] is a comprehensive malware analysis tool which implements data and address tainting techniques of IFT. The given system provides a holistic system-wide investigation in an emulated environment. An informationtrace in the form of a taint graph is presented during the analysis of a monitored program and contains information like network packets payload, function parameters and data accessed during the execution. While a program is emulated, the analysis is performed at the kernel level to intercept the system calls and APIs are invoked. The system designers, however, didn't address the notion of unpacking the binaries in this tool.

*CWSandbox* [28] is a Function Call monitoring tool that supports both user and kernel mode behavior analysis to observe calls made to the APIs and systems calls. A hook function, that implements a monitoring function, is invoked each time when an API is called. A control process is defined which monitors and report the analysis whenever a hook function is called. This enables the analyst to monitor and analyze the child processes that can be launched by a malware. To increase the performance of the analysis process, a custom DLL is compiled before the analysis process in order to have a quick access to the required hook functions. API and systems calls are written into a trace file that provides a top-bottom view of functions calls made during the execution.

*Ether* [29] uses hardware virtualization techniques to monitor functional behavior of a program on individual instruction level. A trap flag is setup when a monitored program is launched. This enables the analyst to debug the program's functional behavior as the trap flag is used as a breakpoint during execution. The presented technique can only retrieve system call information which is implemented using the SYSENTER instruction. The implemented flagging phenomenon also enables tracking and recording of program instructions resulting in reporting of rich information to the analyst detailing individual instruction execution.

*TQana* [30], is an Internet Explorer browser plug-in that analyze the dynamic behavior of a Spyware. TQana can observe the functional behavior as well as the information trace of spyware execution. The analysis performed is emulated at the kernel level where the monitoring program has access to all the calls made. Tainting or IFT is implemented using the Navigate event of the web browser which introduce taints to the system whenever a URL is entered in the address bar, or a link is clicked on a web page. Furthermore, contents accessed by a web browser and bookmarks are also tainted. API and System call monitoring is implemented using the Component Object Model subsystem (COM).

For future attacks, malwares like Rootkits and Spywares hook themselves to system APIs and is activated on desired events. This hooking is done in a stealthy mode to avoid detection. *Hookfinder* [31] is a tool that is used to identify and detect these bogus hooks, presenting the findings in a detailed report. Hookfinder is implemented in emulated environment where analysis is carried out with kernel mode privileges. Hook detection is done by monitoring the control flow of the system processing. The tool observes the data and address taints and whenever the control flow (instruction pointer) is diverted to an unexpected API or system call (or a tainted value in this case), it detects a hook and alerts the analyst. The analysis is done on a single-process level. The analysis report presented reflects the specific API or function to which a hook was attached and a trace of how the instruction pointer was manipulated.

*Norman Sandbox* [32] is a functional call monitoring system which is implemented in a simulated environment where analysis is performed in the user space. It is designed to analyze worms and viruses which use network shares and email as infection vectors. The authors argued that it can be used to analyse packed malware binaries as the infection is already contained in a simulated environment. However, they didn't provide any unpacking technique instead suggested that in order to increase analysis performance, packers can be detected with known databases containing packer signatures. Moreover, it is advised to simulate all network services as some malware intends to use various network ports and services, for instance, SMB 137/138/445 in case of network sharing. If a read or write request to certain service is returned null, the malware might terminate its existence resulting the analysis to fail entirely.

*Justin* [33] is a scheme to handle packed binaries. The theme of this model is that a malware can be detected immediately after the binary is unpacked by its unpacking routine so that it is readily available to be detected by any antimalware software. The scheme can be used as a plugin on the top of an AV which doesn't provide the analysis and detection of un-packer. The external anti-malware software is activated once the Write *xor* Execute condition is detected as true in the memory. The scheme is based on two assumptions:

   i.   Irrespective of the packing techniques used by a malware, the original binary will execute from the same memory address after it is unpacked.
   ii.  After the unpacking, full control is given to the original packed binary.

Logically, these two assumptions remain valid for most of the malwares. However, they can be seen as very strict rules and can be bypassed by malwares which evade detection techniques using multiple layers of packing.

*PolyUnpack* [34] on the other hand is more focused on the un-packer execution. It observes the dynamic behavior of the un-packer execution using both static and dynamic analysis. The packed binary is disassembled and is analysed in a static approach while the packed binary is executed and is observed dynamically at each instruction level. Both the operations are done at the memory level. The packer code is identified by comparing the instruction trace in both the dissembled and the executed binaries. If, in either of the analysis the trace is not matched, it indicates the packed code is about to be unpacked.

## 5. COMPARATIVE ANALYSIS

This section presents an analysis of the summarized tools as per the proposed framework detailed in section 3. The objective of this analysis is to provide the reader a clear idea of how the mentioned tools and techniques incorporate implementation strategies and the support they provide for function call monitoring and IFT analysis. A comparative analysis is shown in the table 5.

It can be seen that most of the tools, Panorama, CWSandbox, TQana, HookFinder, Justin, have focused the kernel level implementation to get a detail understanding of the malware. Malware uses different APIs to call common system calls. They are more interested in system calls to achieve privilege access. APIs provide a medium to connect to them which implies that analysis concentration should be more focused on intercepting and analyzing system calls. This trend can be seen in the studied tools where all of them provide mechanisms to analyze the system calls. IFT is only offered in HookFinder, TQana and Panorama. However, the objective of performing information tainting is different. HookFinder implements IFT to detect the stealthy hooks within a system whereas, TQana and Panorama provide tainting techniques to monitor and analyze sensitive information leakage.

Table 5. Comparative Analysis

| Strategies | Panorama | CW Sandbox | Ether | TQana | Hook finder | Norman | Justin | Poly Unpack |
|---|---|---|---|---|---|---|---|---|
| **Mode** | | | | | | | | |
| User | - | ✓ | - | - | - | ✓ | ✓ | ✓ |
| Kernel | ✓ | ✓ | ✓ | ✓ | ✓ | - | ✓ | - |
| **Environment** | | | | | | | | |
| Emulated | ✓ | - | - | ✓ | ✓ | - | - | - |
| Virtual | - | - | ✓ | - | - | ✓ | - | - |
| **Methods & Support** | | | | | | | | |
| API Hooking | ✓ | ✓ | - | ✓ | ✓ | ✓ | - | - |
| System Calls | ✓ | ✓ | ✓ | ✓ | ✓ | ✓ | - | - |
| Packer Analysis | - | - | - | - | - | - | - | ✓ |
| Packer Reconstruction | - | - | - | - | - | - | - | ✓ |
| W ⊕ X Policy | - | - | - | - | - | - | ✓ | - |
| Trace | ✓ | ✓ | | - | ✓ | - | - | - |
| IFT | ✓ | - | - | - | ✓ | | | |
| **Process Scope** | | | | | | | | |
| Single | - | ✓ | ✓ | ✓ | ✓ | ✓ | ✓ | ✓ |
| Spawned | - | | - | - | - | - | - | - |
| System Wide | ✓ | - | ✓ | ✓ | - | - | - | - |

## 6. CONCLUSION

In this paper, two key techniques of dynamic analysis, IFT and Function Call Monitoring, were detailed. Due to the diversity of concepts and strategies associated with these techniques, it is hard to identify and select a suitable tool. A framework is presented which highlight the key attributes of interest in dynamic malware analysis and provides a general yet detailed enough reference to compare and analyse different analysis tools.

**Waqas Aman** received his bachelor's degree in Computer Sciences from University of Peshawar in 2008 and earned his MS-InfoSec degree from National University of Sciences & Technology Pakistan in 2011. He is currently pursuing his Ph.D. in Information Security at Gjøvik University College Norway, working as a Research Fellow in the Adaptive Security for Smart Internet of Things in eHealth (ASSET) Project financed by the Research Council of Norway. He also have industrial expertise in technical information security as well as in product management and has earned various international certifications including CEH, CEI, ECSA, ENSA, EDRP and ITIL® and remained Certified Instructor for EC-Council™ Technical Certifications in South Asia.

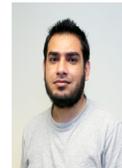